\documentclass[prl,aps,floats,twocolumn,showpacs,superscriptaddress,
preprintnumbers,showkeys]{revtex4-1}

\usepackage[latin1]{inputenc}                    
\usepackage{graphicx}                            
\usepackage{latexsym}                            
\usepackage{amsfonts}                            
\usepackage{amssymb}                             
\usepackage{amsmath}                             
\usepackage[mathscr]{eucal}                      
\usepackage{dcolumn}                             
\usepackage{theorem}                             
\newcommand{\bc}{\begin{center}}
\newcommand{\ec}{\end{center}}
\newcommand{\be}{\begin{equation}}
\newcommand{\ee}{\end{equation}}
\newcommand{\bea}{\begin{eqnarray}}
\newcommand{\eea}{\end{eqnarray}}

\begin{document}

\preprint{
\vbox{
\hbox{ADP-12-19/T786}
}}

\title[Volume Dependence of the Axial Charge of the Nucleon]{Volume
  Dependence of the Axial Charge of the Nucleon}
\author{N.~L.~Hall}\affiliation{ARC Centre of Excellence for Particle Physics 
at the Terascale, School of Chemistry and Physics, 
University of Adelaide,
Adelaide SA 5005, Australia}
\affiliation{CSSM, School of Chemistry and Physics, 
University of Adelaide,
Adelaide SA 5005, Australia}
\author{A.~W.~Thomas}\affiliation{ARC Centre of Excellence for Particle Physics 
at the Terascale, School of Chemistry and Physics, 
University of Adelaide,
Adelaide SA 5005, Australia}
\affiliation{CSSM, School of Chemistry and Physics, 
University of Adelaide,
Adelaide SA 5005, Australia}
\author{R.~D.~Young}\affiliation{ARC Centre of Excellence for Particle Physics 
at the Terascale, School of Chemistry and Physics, 
University of Adelaide,
Adelaide SA 5005, Australia}
\affiliation{CSSM, School of Chemistry and Physics, 
University of Adelaide,
Adelaide SA 5005, Australia}
\author{J.~M.~Zanotti}\affiliation{CSSM, School of Chemistry and Physics, 
University of Adelaide, Adelaide SA 5005, Australia}

\begin{abstract}
It is shown that the strong volume-dependence of the axial charge of the nucleon 
seen
in lattice QCD 
calculations can be understood 
quantitatively in terms of the pion-induced interactions between 
neighbouring nucleons. The associated wave function renormalization leads to 
an increased suppression of the axial charge as the strength of the 
interaction increases, either because of a decrease in lattice size or in 
pion mass.
\end{abstract}



\maketitle

The axial form factor of the nucleon, $G_A(Q^2)$, encodes details of
the nucleon's non-perturbative structure and plays a key role in its
properties under chiral symmetry~\cite{Thomas:2001kw}.  While the
axial charge, $g_A \equiv G_A(0)$, has been known accurately for many
years through measurements of neutron $\beta$-decay, and the shape is
well described by a dipole form with mass parameter around 1 GeV, the
calculation of these properties within QCD still presents significant
challenges.  In particular, the mean square radius of the axial form
factor in modern lattice QCD simulations is almost a factor two too
small compared with experiment~\cite{arXiv:1001.3620} and $g_A$ itself
exhibits a remarkable dependence on the size of the space-time lattice
used~\cite{arXiv:0801.4016,hep-lat/0603028,hep-lat/0510062,hep-lat/0306007,Alexandrou:2010hf,Capitani:2012gj,Pleiter:2011gw}.

The resolution of these challenges is made especially urgent by the
recent contributions of lattice QCD towards unravelling the origin of
the spin of the
proton~\cite{arXiv:1001.3620,arXiv:0705.4295,arXiv:0803.2775,arXiv:0911.1974}.
{}For example, the isovector combination of the orbital angular
momentum carried by the quarks, $L^u - L^d$, is dramatically different
in several widely used
models~\cite{arXiv:0908.0972,hep-ph/0502030,arXiv:0709.4067} and it
can be deduced from a lattice calculation of $J^u - J^d$ by
subtracting the value of $g_A$ implicit in that same
calculation. While such estimates have already been published, one
cannot assess their reliability without a better understanding of the
systematic error associated with the finite lattice
volume~\cite{EPJ_A46}.

In this Letter we explain the origin of the dramatic volume dependence
of $g_A$ in terms of the pion exchange force between neighbouring
nucleons on the periodic lattice used in lattice QCD simulations. 
This quantitative, model independent explanation provides a framework
within which to fully assess the systematic errors associated with the
determination of the quark spin and orbital angular momentum as
discussed above.

One of the earliest attempts to understand why $g_A$ might decrease rapidly 
at low pion mass as the lattice size decreases relied on the fact that the 
axial current has a contribution proportional to $\vec{\nabla} \phi$, with 
$\phi$ the pion field. 
In a finite box one may expect that the axial charge, given 
as the volume integral of the matrix element of the axial current would receive 
a contribution from a surface integral of the pion field, 
arising through Green's Theorem~\cite{hep-ph/0108015}.
Promising as this idea seems, it has been 
proven that with the periodic boundary condition imposed on the quark fields, 
this surface contribution must vanish and hence cannot be the source 
of the problem~\cite{hep-lat/0112014}.

Next one might think of the volume dependence of the chiral loops which 
renormalize the axial charge~\cite{hep-lat/0206001}, 
with wave function renormalization reducing 
$g_A$ and the vertex renormalization, especially that associated with the 
axial current acting on an intermediate $\Delta$ baryon, 
tending to increase it~\cite{TRI-PP-82-29,TRI-PP-81-16}.
These two contributions tend to effectively 
cancel each other over a wide range of bag radii, 
so that the renormalized and un-renormalized axial charges 
are very close to each other. Given that the volume dependence of the wave 
function and vertex renormalization is very similar for pions 
which do not travel outside the lattice volume, this is also not the 
source of the rapid variation of $g_A$. Similar conclusions have been reached 
within other approaches~\cite{Beane:2004rf,arXiv:1108.5594}.

What has not previously been recognised, however, is the effect
associated with the fact that on a finite lattice each nucleon must
interact with its neighbours through pion exchange.  (Alternatively,
the pion travels outside the lattice volume or ``round the world''.)
In terms of its effect on the axial charge this may be treated in an
analogous way to the renormalization of an individual nucleon.
Once again the wave function renormalization, that is the probability
to find a bare nucleon rather than one which has emitted a pion that
will be absorbed by one of its neighbours, reduces $g_A$, while the
vertex renormalization acts to increase it.
In this case, however, the $N \rightarrow \Delta \pi$ transition is
suppressed by an extra factor of $\exp[ - \delta L]$, with $\delta =
m_\Delta - m_N$ and $L$ the lattice size, which is the same as the
separation between nearest neighbours on the lattice.
As a result, at all but the very smallest lattices (where the
neighbouring nucleons would overlap anyway), this contribution is not
effective at countering the effect of wave function renormalization.
Similarly, the vertex renormalization involving $N \pi$
intermediate states is small for a single nucleon (namely that the
interchange of $\sigma_z \tau_3$ with the $\sigma \tau$ factors in the
pion loop generates a factor of 1/9 compared with the wave function
renormalization).
Hence the wave function renormalization dominates and $g_A$ is
suppressed as the factor $\exp[- m_\pi L]$ increases, or $m_\pi L$
decreases.

\begin{figure}
\begin{center}
\includegraphics[width=3.3in]{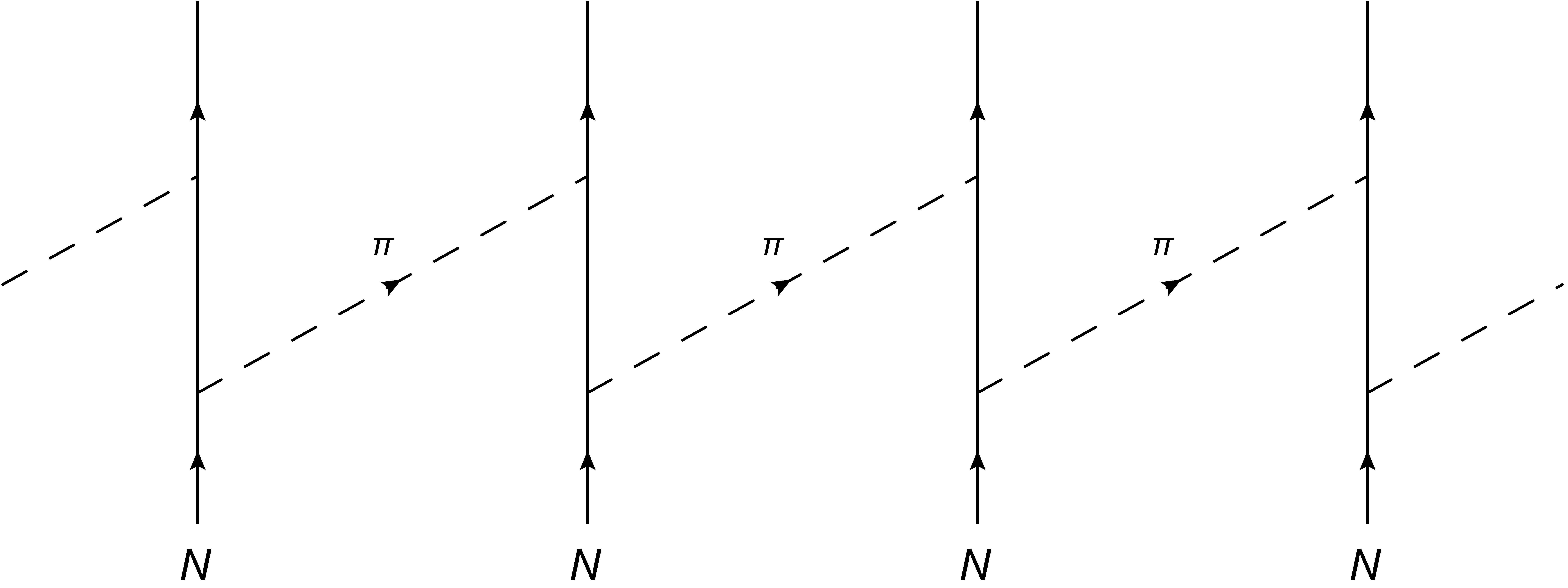}\\
\caption{Schematic illustration of pion emission and absorption on a
  periodic lattice.}
\label{PerLat}
\end{center}
\end{figure}

\begin{figure}
\begin{center}
 \includegraphics[width=2.7in]{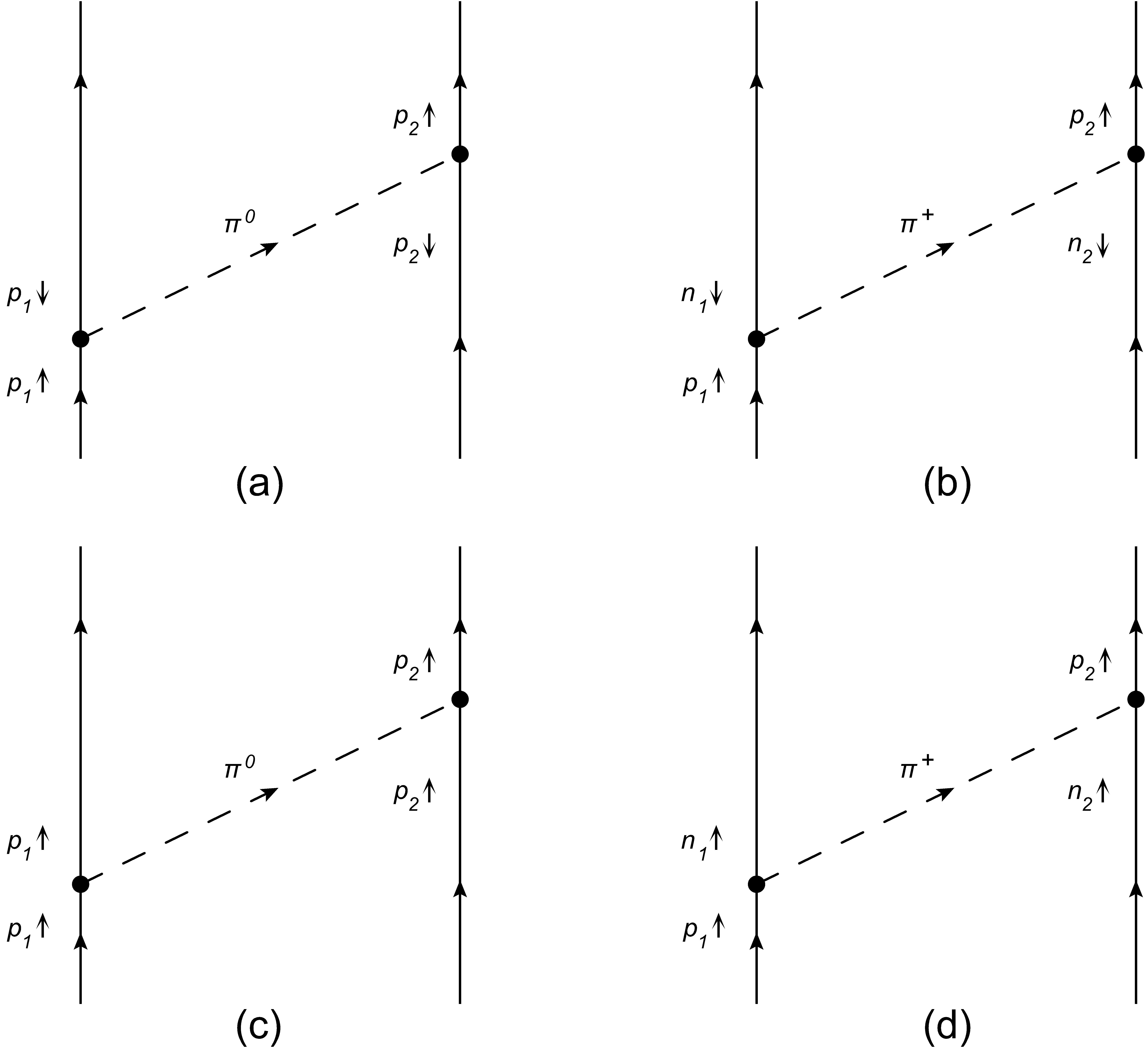}\\
 \caption{Diagrams which contribute to the wave function renormalization.}
 \label{4Diag}
\end{center}
\end{figure}

The emission of a pion which is absorbed on a neighbouring nucleon is
illustrated schematically in Fig.~\ref{PerLat} for a nucleon on a
periodic lattice.
Indeed, when the nucleon under consideration emits a pion, every copy
also emits a meson and the particular one in the box under
consideration must absorb a pion emitted by the other neighbour.
Looking at the contribution to the wave function renormalization
coming from a {\it single} pair of neighbouring nucleons -- as in
Fig.~\ref{4Diag} -- one sees (in the heavy baryon approximation which
should be perfectly adequate for this purpose) that
\begin{eqnarray}
\delta Z_2 & = &  \bigg\langle N_{1\alpha'} N_{2\beta'} \; 
\bigg| \, \frac{\tau_{2i} \tau_{1i}}{16 \pi^3}
 \left( \frac{g_A}{2 \, f_{\pi}} \right)^2
\nonumber \\
&&  \hspace{-1.0cm} \times
\vec{\sigma}_2 \cdot \frac{\nabla}{i} \vec{\sigma}_1 \cdot \frac{\nabla}{i} 
\int d^3 k \frac{ e^{i \vec{k}
       \cdot \vec{L}}}
 {(k^2 + m_\pi^2)^{3/2}} \bigg|   N_{1 \alpha} \, N_{2 \beta} \bigg\rangle \, ,
\label{eq:dZ21}
\end{eqnarray}
where $\alpha$, $\beta$, $\alpha'$ and $\beta'$ represent the nucleon spin
states.

We suppose that the proton spin points up, along the $z$-axis, and
consider first the interaction through the exchange of a $\pi^0$ with
the nearest neighbour along the $y$-axis.
This is illustrated by Fig.~\ref{4Diag} (a), where the pion emission
necessarily flips the spin of the proton.  In this case $\langle
\tau_{2i} \tau_{1i} \rangle = +1$ and the contribution to the wave
function renormalization becomes
\begin{eqnarray}
\delta Z_2^{(a)} &=&
\frac{1}{4 \pi^2} \left( \frac{g_A}{2 f_{\pi}} \right)^2 
\langle N_{1 \downarrow} N_{2 \uparrow} | \sigma_{2y} \sigma_{1y} |
N_{1 \uparrow} N_{2 \downarrow} \rangle \nonumber \\
&&  \hspace{-1.3cm} \times \frac{\partial^2}{\partial L^2} 
\int_0^\infty dk k \frac{\sin(kL)}{ L \, (k^2 \, + \, m_\pi^2 )^{3/2}} \, .
\label{eq:dZ2a}
\end{eqnarray}
Defining the usual spin raising and lowering operators
\begin{equation}
\sigma_{\pm} =  \frac{1}{2} (\sigma_x \pm i \sigma_y ) ; 
\qquad \sigma_0 = \sigma_z \, ,
\end{equation}
we see that 
\begin{equation}
\sigma_{1y} \, \sigma_{2y} = 2 \, (\sigma_{1+} \, \sigma_{2-} + 
\sigma_{1-} \, \sigma_{2+}) \, ,
\end{equation}
so the spin matrix element gives a factor of 2 in this case. Finally, 
evaluating the momentum integral we find
\begin{eqnarray}
\delta Z_2^{(a)} &=&
\frac{1}{2 \pi^2} \left( \frac{g_A}{2 f_{\pi}} \right)^2
\frac{\partial^2}{\partial L^2} K_0(m_\pi L) \, \nonumber \\
&=&
\frac{m_\pi^2}{4 \pi^2} \left( \frac{g_A}{2 f_{\pi}} \right)^2
(K_0(m_\pi L) + K_2(m_\pi L) ) \, ,
\label{eq:dZ2aK02}
\end{eqnarray}
where $K_n$ are the modified Bessel functions (e.g., $K_0(z) \rightarrow 
(\pi/2)^{1/2} \exp(-z)$ as $z \rightarrow \infty$).

Because of the periodicity of the lattice one can also exchange a
$\pi^+$ with the neighbouring nucleon along the $y$-axis.
In this case the isospin factor is two, rather than one, and hence
$\delta Z_2^{(b)} = 2 \delta Z_2^{(a)}$.
For the neighbouring nucleon along the $z$-axis the spin factor is
unity.
That is, the spin is not flipped. Thus $\delta Z_2^{(c)}$ and $\delta
Z_2^{(d)}$ are each half as large as $\delta Z_2^{(a)}$ and $\delta
Z_2^{(b)}$, respectively.
This asymmetry between lattice sites perpendicular and parallel to the
nucleon spin is a natural consequence of the tensor force generated by
the exchange of a pseudoscalar meson.


The expressions above give the contribution to the wave function
renormalization associated with pion exchange between a single pair of
nucleons.
To ensure that we represent what goes on on the lattice we need to
include {\it all} neighbours which make significant contributions.
There are four nearest neighbours in the $x-y$ plane and two along the
$z$ direction.
Thus the total contribution to the wave function renormalization from
nearest neighbours, $\delta Z_2^{\rm nn}$, is
\begin{eqnarray} 
\delta Z_2^{\rm nn} &=& 
15 \frac{m_\pi^2}{4 \pi^2} \left( \frac{g_A}{2 f_{\pi}} \right)^2
(K_0(m_\pi L) + K_2(m_\pi L) ) \, , \nonumber \\
&\equiv& 15 {\cal F} (m_\pi^2,m_\pi L) \, . 
\label{eq:dZ2nn}
\end{eqnarray}

As explained earlier, for $g_A$ the vertex renormalization is not very
effective in countering the wave function renormalization when the
$\Delta$ excitation is suppressed.
However, for a lattice which is not spherically symmetric the
suppression is not simply 1/9 but differs from nearest (n1) to
next-to-nearest (n2) to next-to-next-to-nearest neighbours (n3). 
It is a trivial calculation to show that for n1 the vertex
renormalization is one fifth of $\delta Z_2^{\rm nn}$. Thus the total
relative correction to $g_A$ (i.e. $g_A(1-\delta g_A)$) from nearest
neighbours is:
\begin{equation}
\delta g_A^{\rm n1} = \frac{4}{5} \delta Z_2^{\rm nn} 
\label{eq:dgAnn}
\end{equation}
\begin{figure}[ht]
\begin{center}
 \includegraphics[width=0.95\columnwidth]{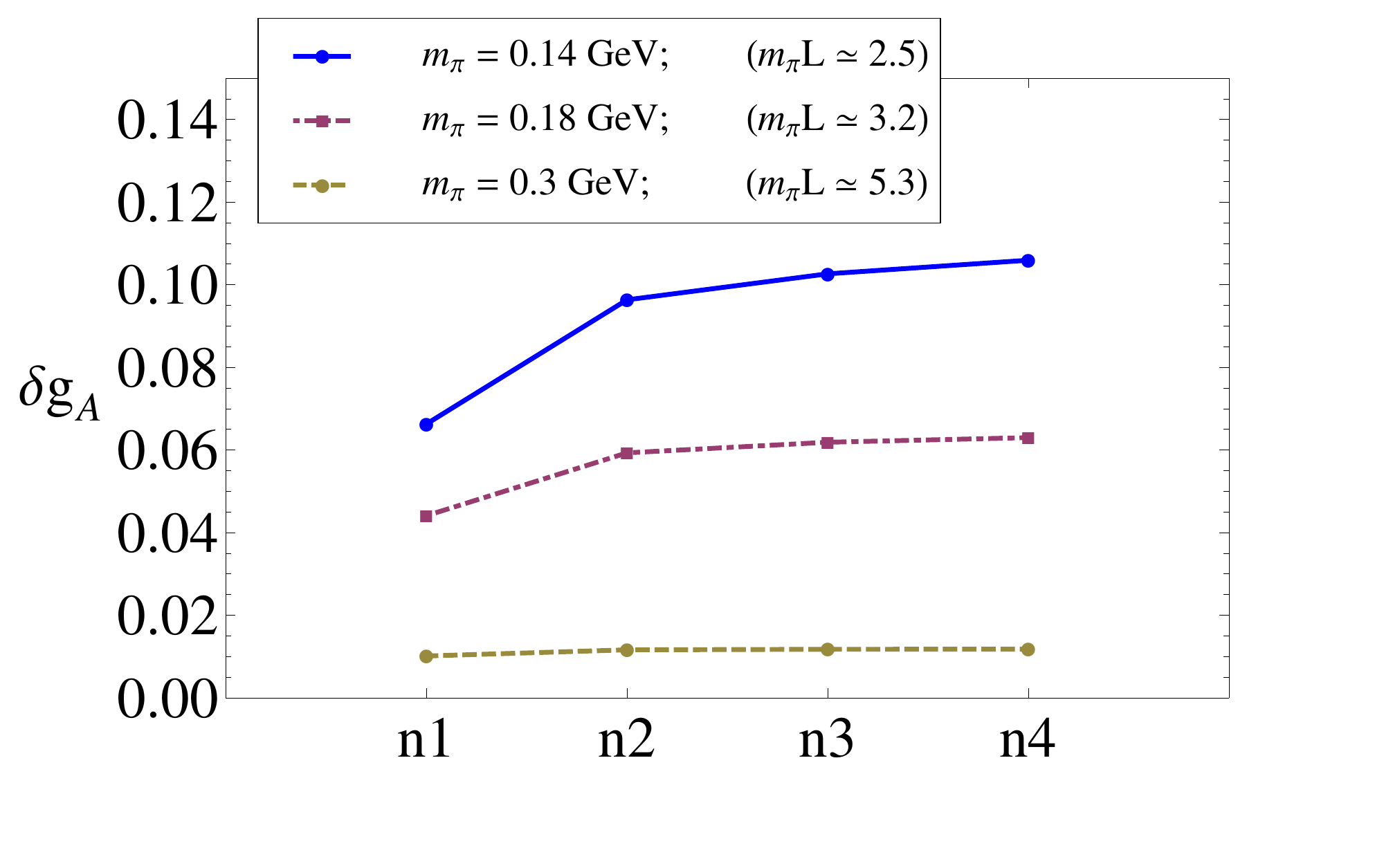}\\
 \caption{Illustration of the convergence of the correction to $g_A$
   calculated on a lattice 3.5 fm on a side as we include the nearest
   neighbour (n1) up to next-to-next-to-next-to-nearest neighbour
   (n4), for three values of $m_\pi L$.}
\label{Con35}
\end{center}
\end{figure}
Turning to the effect of the next-to-nearest neighbours, we note that 
there are more of them (12 instead of 6) but that their distance from 
the nucleon under study is $\sqrt{2} L$. 
Thus, provided $m_\pi L$ is relatively large, we expect that the
correction to $g_A$ should converge relatively quickly. 
In the present work we consider the corrections out to and including
next-to-next-to-nearest neighbours, or a distance $\sqrt{3} L$.
In these cases, the factor one fifth for n1 becomes one sixth for n2 and
one ninth for n3. In summary, the total corrections to $g_A$ from
next-to-nearest and next-to-next-to-nearest neighbours are
\begin{align*}
\delta g_A^{\rm n2} &= \frac{5}{6} 24 {\cal F} (m_\pi^2,\sqrt{2} m_\pi L) \, ,\\
\delta g_A^{\rm n3} &= \frac{8}{9} 12 {\cal F} (m_\pi^2,\sqrt{3} m_\pi L)\, .
\label{eq:dgAn2n3}
\end{align*}
The degree of convergence for three relatively light quark masses is
illustrated in Fig.~\ref{Con35} for a 3.5 fm lattice. 
However we point out that the convergence of the correction is
solely dependent on $m_\pi L$, and the figure displays the convergence
down to $m_\pi L\simeq 2.5$.
Here we also show the contribution from the next $(2L)$ term which
justifies our truncation at a distance of $\sqrt{3} L$ in the
remainder of this letter.
\begin{figure}[ht]
\begin{center}
 \includegraphics[width=3.2in]{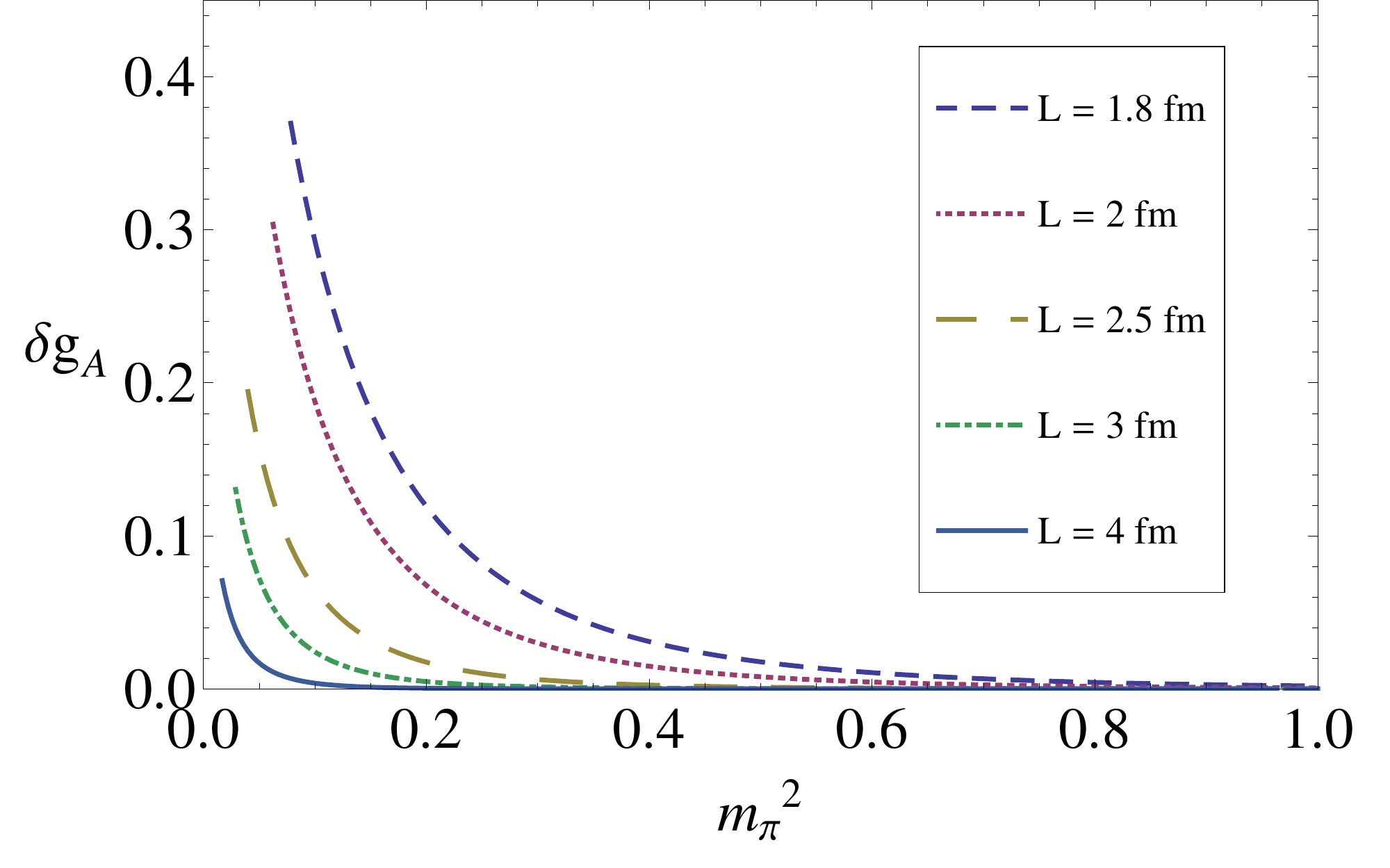}\\
 \caption{The total reduction in $g_A$ from wave function and vertex 
renormalization is illustrated as a function of pion mass (in GeV) 
{} for several lattice sizes.}
\label{NNLg}
\end{center}
\end{figure}
The total reduction in $g_A$ to this order is illustrated in
Fig.~\ref{NNLg} --- note that for each volume, the curves are plotted
only down to a minimum reliable pion mass estimated by $m_\pi L= 2.5$.
This figure makes it clear that these corrections are substantial. Modern lattice
simulations of $g_A$ are now available at masses as low as 0.3 GeV,
but even there the correction is almost 20\% on a 2 fm box and 6\% at
2.5 fm.  Even more significant, we see that as the pion mass
approaches the physical value the correction is as large as 10\% on a
4 fm lattice.  This makes very clear the challenge of a brute force
determination of this quantity at the physical quark mass.

\begin{figure}[ht]
 \begin{center}
 \includegraphics[width=3.2in]{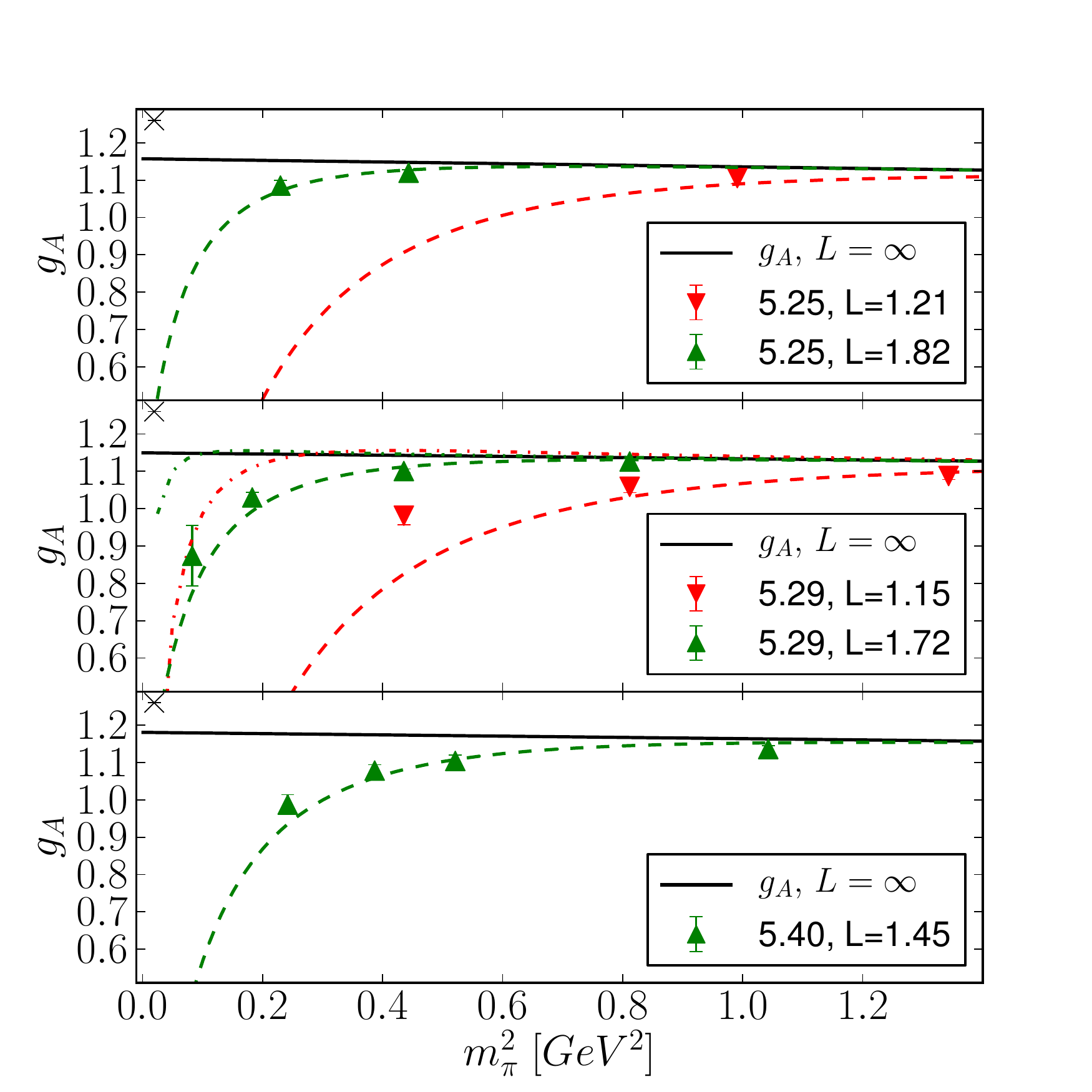}\\
\caption{The axial charge $g_A$, with the finite volume correction
  associated with pion exchange between neighbouring nucleons
  included, is plotted for several different values of the lattice
  size, $L$ (in fm).  Also plotted are lattice calculations of $g_A$
  from QCDSF~\cite{hep-lat/0603028,Pleiter:2011gw}
  with three different $\beta$'s (lattice spacings),
  $\beta=5.25\,(a=0.076\,\mathrm{fm})\ \beta=5.29\,(a=0.072\,\mathrm{fm})\ \beta=5.40\,(a=0.060\,\mathrm{fm})$,
  on multiple volumes.  For comparison, in the middle figure we also
  show (dot-dash line) the predicted finite size effects from
  \cite{hep-lat/0603028}, using the SU(6) $\Delta$ couplings \cite{Beane:2004rf}.}
 \label{QCDSF-gA}
 \end{center}
\end{figure}

\begin{figure}[ht]
 \begin{center}
 \includegraphics[width=3.2in]{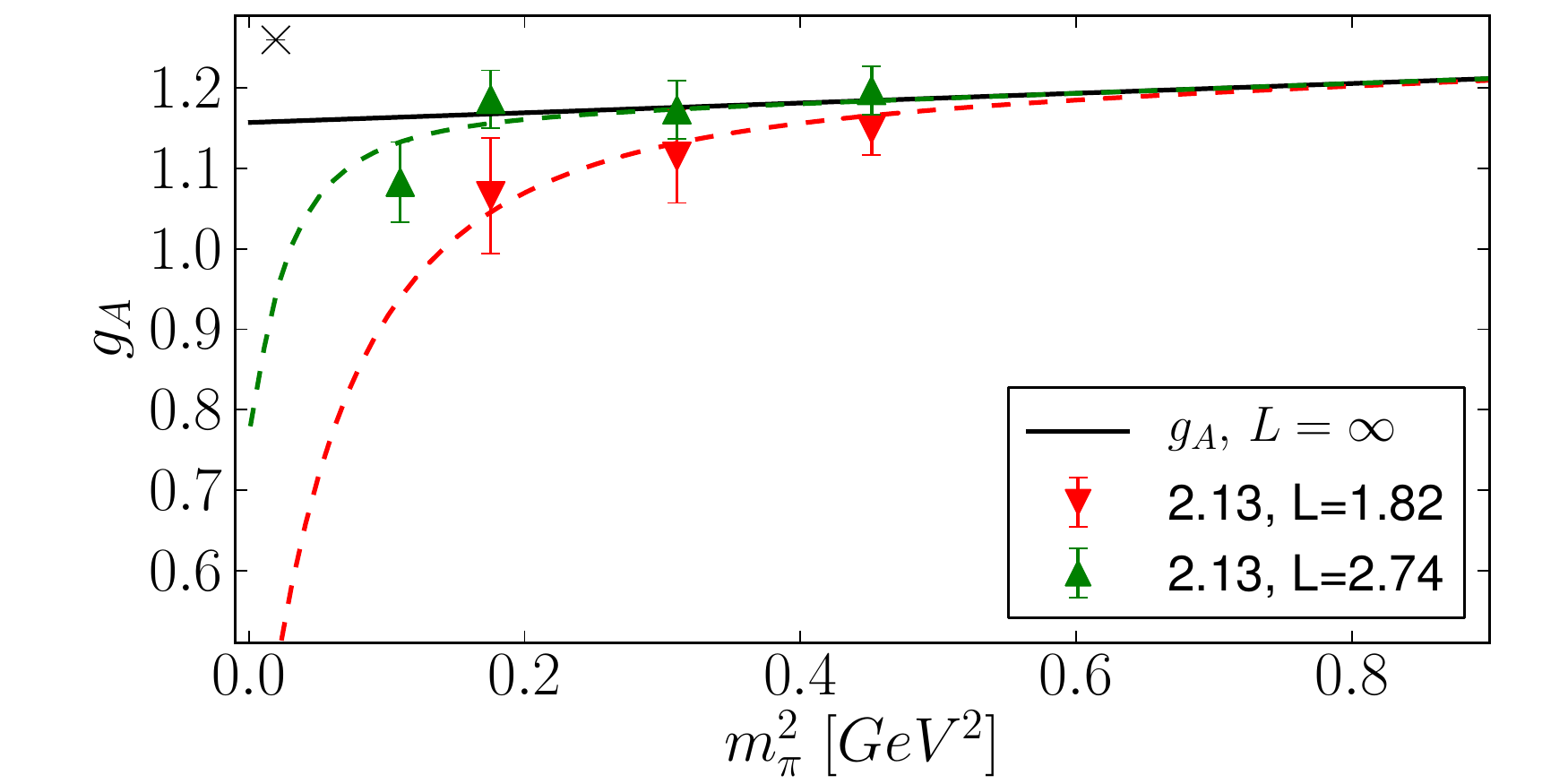}\\
\caption{
Curves as described in Fig.~\ref{QCDSF-gA}, plotted with lattice calculations of $g_A$ 
by Yamazaki {\it et al.}~\cite{arXiv:0801.4016}.}
 \label{Nexdifg}
 \end{center}
\end{figure}

We show the effect of this correction applied to a simple fit linear
in $m_\pi^2$, $g_A^0(1-\delta g_A) + Bm_\pi^2$, with two fit
parameters $(g_A^0,B)$ applied to recent lattice data from the QCDSF
collaboration in Fig.~\ref{QCDSF-gA} and the RBC collaboration in
Fig.~\ref{Nexdifg}.
Note that we have made no attempt here to include any non-analytic
effects in the pion mass dependence of $g_A$, although in this case,
these effects are anticipated to be small
\cite{hep-lat/0603028,hep-lat/0206001}.
The very dramatic reductions in $g_A$ at small lattice sizes and low
pion mass are very clearly seen there, and we find remarkable
agreement between the present calculation and those data.

Earlier work has suggested that similarly large finite-volume effects
can be described by the discretisation of standard one-loop
corrections \cite{hep-lat/0603028}.
The enhancement seen there can by understood noting that both the input
$g_{\Delta N}$ coupling and the fitted $g_{\Delta\Delta}$ coupling are
about 1.5 times larger than those predicted by SU(6) relations. 
In the second panel of Fig.~\ref{QCDSF-gA}, the standard contributions
using SU(6) estimates for the couplings are shown for the two relevant
volumes. We see that the effects are insignificant compared to the
exchange effects investigated in the present work.

The calculations reported here show clearly that a significant
part of the hitherto unexplained quenching of $g_A$
on small lattice volumes originates with the pion exchange
force between neighbouring nucleons on a periodic lattice. 
In retrospect, this explanation seems to be very natural, although the
key role of wave function renormalization was somewhat unexpected.

With this fascinating issue resolved, one may ask what other problems 
being addressed by lattice QCD calculations may be similarly affected. 
The study of the fraction of the nucleon spin carried by its quarks is
certainly a topical example, as is the orbital angular momentum of the
quarks and gluons and hadron magnetic moments.  
The tensor nature of the pion exchange force suggests that special
care may be required in those cases where spin related deformation is
important \cite{arXiv:1011.3233}, for example in the study of the
E2/M1 ratio for the $\Delta$.

%
\section*{Acknowledgements}
%
%
It is a pleasure to acknowledge numerous valuable discussions with
P.~A.~M.~Guichon at early stages of this investigation.  This work was
supported by the Australian Research Council through the ARC Centre of
Excellence for Particle Physics at the Terascale and through grants
FL0992247 (AWT), DP110101265 (RDY) and FT100100005 (JMZ), as well as
by the University of Adelaide.
We would also like to thank the QCDSF Collaboration for giving us
access to some of the unpublished lattice results appearing in
Fig.~\ref{QCDSF-gA}.
%
%


\end{document}